\documentclass[runningheads]{llncs}

\usepackage{xcolor}
\usepackage{afterpage}
\usepackage{graphicx}
\usepackage{amsmath,amssymb,amsfonts}
\usepackage{bm}
\usepackage{braket}
\usepackage{float}
\usepackage{algorithm}
\usepackage{algpseudocode}
\usepackage{booktabs}
\usepackage[hidelinks]{hyperref}
\usepackage[capitalise]{cleveref} %
\usepackage{color}
\usepackage[T1]{fontenc}
\usepackage[acronym]{glossaries}
\usepackage[inline]{enumitem}
\usepackage[numbers,sort]{natbib}
\usepackage[skip=1pt]{caption}
\usepackage{orcidlink}

\usepackage[acronym]{glossaries}
\makeglossaries

\newcommand{\ourparagraph}[1]{\smallskip\noindent\emph{#1}}

\renewcommand{\refname}{References}   
   


\begin{document}

\title{Multi-Step Semantic Reasoning in \\Generative Retrieval}

\titlerunning{Multi-Step Reasoning for GR}

\author{Steven Dong\orcidlink{0009-0007-6129-921X}
 \and
Yubao Tang\orcidlink{0009-0003-8010-3404} \and
Maarten de Rijke\orcidlink{0000-0002-1086-0202}}
\authorrunning{S. Dong et al.}
%
\institute{University of Amsterdam, Amsterdam, The Netherlands
\\
\email{steven.dong@student.uva.nl, }
\email{\{y.tang3, m.derijke\}@uva.nl} 
}

\maketitle          

\begin{abstract}

Generative retrieval (GR) models encode a corpus within model parameters and generate relevant document identifiers directly for a given query. While this paradigm shows promise in retrieval tasks, existing GR models struggle with complex queries in numerical contexts, such as those involving semantic reasoning over financial reports, due to limited reasoning capabilities. This limitation leads to suboptimal retrieval accuracy and hinders practical applicability.
We propose ReasonGR, a framework designed to enhance multi-step semantic reasoning in numerical contexts within GR. ReasonGR employs a structured prompting strategy combining task-specific instructions with stepwise reasoning guidance to better address complex retrieval queries. Additionally, it integrates a reasoning-focused adaptation module to improve the learning of reasoning-related parameters.
Experiments on the FinQA dataset, which contains financial queries over complex documents, demonstrate that ReasonGR improves retrieval accuracy and consistency, indicating its potential for advancing GR models in reasoning-intensive retrieval scenarios.

\end{abstract}

\section{Introduction}

Generative retrieval (GR) has emerged as a promising paradigm in information retrieval. Unlike traditional dense retrieval methods that rely on explicit index-retrieve pipelines \cite{karpukhin2020dense,tang2021improving,gao2022unsupervised}, GR models learn to encode the entire corpus within their parameters and directly generate relevant document identifiers (docids) in response to queries \cite{tay2022transformer}. This approach reduces memory storage requirements and enables more integrated, end-to-end retrieval.

While GR has demonstrated notable results in tasks such as question answering and fact verification \cite{chen-2023-unified,chen2022corpusbrain,chen2022gere,corpusLM}, it still faces challenges in retrieval tasks that require complex inference or aggregation of dispersed evidence, exemplified by financial queries over intricate reports in the FinQA dataset \cite{chen2021finqa}. We refer to this challenge as multi-step semantic reasoning in numerical contexts, which focuses on interpreting relevant numerical information in complex documents \cite{ginting2024seek}.

Current GR methods struggle to effectively address these tasks, mainly due to two factors: 
(i) GR models often treat retrieval as a black-box operation, encoding it implicitly in a single forward pass without explicit, interpretable reasoning steps \cite{tang2023recent}.  
(ii) Most models are trained with maximum likelihood estimation (MLE) \cite{hao2022teacher} on surface-level query-document pairs, which does not explicitly foster multi-step or logical reasoning capabilities \cite{liu2024generative,bose2022controllable}.  

To address these challenges, we propose ReasonGR, a framework designed to enhance multi-step semantic reasoning capabilities of GR models in numerical contexts. ReasonGR employs structured prompting that combines task-specific instructions with chain-structured reasoning prompts to guide step-by-step semantic inference. Moreover, we introduce reasoning adapter modules that enable the model to better capture reasoning-related parameters tailored to the task.

We evaluate ReasonGR on the FinQA dataset \cite{chen2021finqa}, which contains financial queries over complex documents. Our experiments show that ReasonGR improves retrieval accuracy over baseline methods. We also report that the introduced adaptation modules yield efficiency benefits in training.

\section{Related Work} \label{ch:related}

\emph{Traditional information retrieval (IR).}  
Traditional IR systems typically follow a ``index-retrieve’’ pipeline \cite{nogueira2019passage}. Early approaches such as BM25 \cite{robertson2009probabilistic} rely on exact term matching, while dense retrieval methods \cite{karpukhin2020dense} use neural networks to capture semantic similarity by encoding queries and documents into dense vectors.

\ourparagraph{Generative retrieval}  
GR learns to index the corpus and then directly generate docids given a query \cite{tay2022transformer,zhuang2022bridging}. This paradigm internalizes corpus information in model parameters, enabling end-to-end retrieval without an external index. A number of related studies have advanced this line of work \cite{doct5query,zhuang2022bridging}. While these methods achieve promising retrieval performance, their ability to perform complex reasoning, particularly multi-step semantic reasoning required by tasks such as financial document retrieval, remains limited \cite{jiang2022understanding}. Addressing this reasoning gap is essential for improving the applicability of GR in such domains.

\section{Methodology} \label{ch:method}

\begin{figure}[t]
    \centering
    \includegraphics[width=\linewidth]{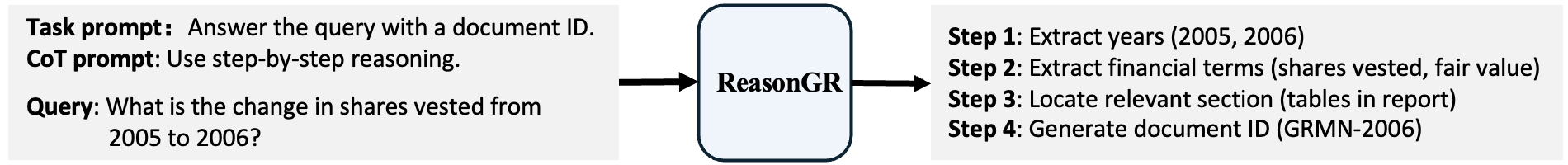}
\caption{ReasonGR performing multi-step semantic reasoning on a FinQA query. The model extracts key info and locates relevant report sections to generate the docid, formed by the company name and report year.}\label{fig:architecture}
\end{figure}

\subsection{Task definition}  
Given a corpus $\mathcal{D} = \{d_1, \dots, d_N\}$ with size $N$, and a set of queries $\mathcal{Q} = \{q_1, \dots, q_M\}$ with size $M$, each document $d_i$ is associated with a unique docid $j_i$, forming the set $\mathcal{J}$. The task is to generate the relevant docid $j_i$ for a given query $q_i$ using a generative model $P_\theta$, which models the conditional probability  
, $P_\theta(j_i | q_i) = \prod_{p=0}^{|j_i|} P_\theta(j_i^p | q_i, j_i^{<p}),$
where $j_i^p$ is the $p$-th token of the docid sequence $j_i$, and $j_i^{<p}$ denotes all previously generated tokens.

\subsection{Model architecture}  
Our model adopts a transformer-based encoder-decoder architecture, following the GR paradigm \cite{tay2022transformer}. 
The backbone encodes the input query and autoregressively decodes the corresponding docid.

To learn reasoning-specific capabilities, we introduce Low-Rank Adaptation (LoRA) modules \cite{hu2022lora} into the backbone. LoRA freezes the original model weights $\theta$ and injects trainable low-rank matrices $\Delta \theta$ to adapt the model to the reasoning-intensive retrieval task: $\theta' = \theta + \Delta \theta, \quad \text{where} \quad \Delta \theta = A B^\top, \quad A \in \mathbb{R}^{d \times r}, B \in \mathbb{R}^{k \times r},$
with rank $r \ll \min(d,k)$. This decomposition allows efficient learning of new parameters dedicated to reasoning, while keeping the number of trainable parameters small.
Furthermore, we apply QLoRA \cite{dettmers2023qlora} to quantize the frozen backbone weights to 4-bit precision, reducing memory and computational costs with minimal performance loss.
LoRA and QLoRA are used jointly in our framework.

\subsection{Reasoning-guided training framework}  
Beyond naive optimization with MLE \cite{tay2022transformer}, our framework improves the model’s reasoning ability through both input construction and training objectives.

\paragraph{Reasoning-enhanced prompting}  
To help the model to structure its generation in a way that encourages multi-step logical inference,
we design our prompts by combining two sets of templates: task-specific instructions and Chain-of-Thought (CoT) additions, as shown in Figure~\ref{fig:architecture}. During training, one of five task templates is randomly selected and combined with one of five CoT instructions to form diverse and informative prompts. 
The five task templates are:  
\begin{quote}
\textit{Answer the query with a document ID.} \\
\textit{Generate the document ID that answers the question.} \\
\textit{Based on the question, predict the document ID.} \\
\textit{Retrieve a document ID that fits the query.} \\
\textit{Using the question, find the document ID.}
\end{quote}

\noindent
The CoT additions encourage stepwise reasoning, examples of which include:  
\begin{quote}
\textit{Use step-by-step reasoning.} \\
\textit{You need to explain your answer.} \\
\textit{Think this through carefully.} \\
\textit{Let's think step-by-step.} \\
\textit{Explain your reasoning before answering.}
\end{quote}

\noindent
These instructions are appended to the task templates to form full prompts.
For CoT prompts, the expected output includes intermediate reasoning steps before the final answer.

\paragraph{Training tasks}
To train the retrieval model, we have two tasks:
\begin{enumerate*}[label=(\roman*)]
\item To memorize the corpus, we learn the mapping from documents to docids using the MLE loss \cite{tay2022transformer,chen-2023-unified}. Since these documents often contain tables with numerical data, which are crucial for retrieval generation, we also incorporate learning from the tables by splitting them and attaching each entry with its corresponding column and row headers.
\item To learn relevance, the model takes queries as input and generates correct multi-step reasoning traces along with the corresponding docids as output. To better approximate the distribution of real-world queries, training data is augmented with multiple pseudo-queries generated per document using a query generation model \cite{zhuang2022bridging}.

\end{enumerate*}

\paragraph{Adaptive penalty scaling loss}  
We design a custom loss function that provides fine-grained supervision by incorporating multiple evaluation signals, including exact match (EM), partial match (PM), set match (SM), and structural similarity (S-Score). The penalty factor scales the cross-entropy loss to emphasize errors on incorrect or incomplete docid predictions proportionally,
$\text{Loss}(\hat{y}, y) = \text{CE}(\hat{y}, y) \times \text{P}_{\text{loss}}(\hat{y}, y),$
where $\text{P}_{\text{loss}}$ is a weighted combination of penalties derived from these metrics. $y$ and $\hat{y}$ are the  model’s predicted sequence, and corresponding ground truth sequence, respectively.

\section{Experimental Setup} \label{ch:setup}

\paragraph{Dataset and metrics}  
We conduct experiments on the FinQA dataset \cite{chen2021finqa}, which contains financial queries requiring complex multi-step numerical reasoning over financial reports and associated tabular data.
We use each document’s filename as a unique docid. 
The dataset includes 8,281 samples, partitioned into training, validation, and test sets with splits of 75\%, 10\%, and 15\%, respectively.
Performance is evaluated using several standard metrics:  
(i) Exact Match accuracy (EM), measuring complete sequence correctness;  
(ii) Part Match accuracy (PM), accounting for ordered token-level correctness;  
(iii) Set Match accuracy (SM), assessing unordered token overlap;  
(iv) Structure Score (S-Score), quantifying relative length differences between predicted and target sequences.
We compare against two representative retrieval methods: BM25 \cite{bm25s}, a sparse retrieval approach, and DSI \cite{tay2022transformer}, a GR method.

\paragraph{Model variants}  
We evaluate three models: our main method, ReasonGR, and two variants. 
\begin{enumerate*}[label=(\roman*)]
    \item ReasonGR combines few-shot prompting with 2 examples, CoT prompting with 2 examples, LoRA adapters, and 4-bit quantization applied on a frozen backbone. 
    \item ReasonGR (Zero) removes prompt training. 
    \item ReasonGR (CoT) extends ReasonGR by modifying the outputs to reflect CoT prompting.
\end{enumerate*}

\paragraph{Implementation details}  
Our backbone model is FLAN-T5 base \cite{chung2024scaling}. 
Docids are formed by concatenating keywords extracted via KeyBERT \cite{grootendorst2020keybert}, with company and year prepended, joined by hyphens.
For each document, we adopt 10 pseudo-queries generated by docT5query \cite{nogueira2019doc2query}.
LoRA weights are initialized with OLoRA \cite{buyukakyuz2024olora} and optimized with LoRA+ \cite{hayou2024lora+} based on an 8-bit AdamW optimizer \cite{dettmers2022optimizers}.
Training uses a batch size of 32 for 100 epochs with early stopping based on PM improvement.  
All experiments are performed on a single NVIDIA H100 GPU.

\section{Experimental Results} \label{ch:results}

\subsection{Main results}
\begin{table}[t]
\caption{Retrieval performance on evaluation and test sets. Bold indicates best performance. PM and SM are not applicable to BM25.}
\label{tab:all_results}
\centering
\begin{tabular}{l ccc ccc}
\toprule
\textbf{Model} & \multicolumn{3}{c}{\textbf{Evaluation}} & \multicolumn{3}{c}{\textbf{Test}} \\
\cmidrule(r){2-4}
\cmidrule{5-7}
               & EM & PM & SM & EM & PM & SM \\
\midrule
BM25                 & \textbf{0.623}  & --     & --     & 0.625  & --     & --     \\
DSI & 0.563 & 0.646 & 0.651 & 0.578 & 0.654 & 0.659\\
ReasonGR (Zero)       & 0.572  & 0.732  & 0.748  & 0.601  & 0.750  & 0.767  \\
ReasonGR (CoT)        & 0.571  & 0.728  & 0.748  & 0.612  & 0.755  & 0.774  \\
ReasonGR              & 0.607  & \textbf{0.751}  & \textbf{0.765} & \textbf{0.626}  & \textbf{0.762}  & \textbf{0.779} \\
\bottomrule
\end{tabular}
\vspace{-4mm}
\end{table}

Table~\ref{tab:all_results} presents retrieval performance of our ReasonGR variants alongside traditional baselines BM25 and DSI on both evaluation and test sets. We observe the following:
\begin{enumerate*}[label=(\roman*)]
\item ReasonGR variants consistently outperform DSI across all metrics, indicating the benefit of reasoning-enhanced prompting and parameter-efficient adaptation over the vanilla GR approach.

\item Compared to DSI, ReasonGR models show improvements in PM and SM metrics, reflecting better token-level and unordered component matching. The full ReasonGR model achieves the highest scores on PM and SM, and also improves EM relative to BM25, showing increased accuracy in generating exact docids.

\item Among ReasonGR variants, prompt training proves important: ReasonGR (Zero), without prompt training, performs worse than the full ReasonGR model, confirming the effectiveness of the combined prompting strategies.

\item ReasonGR (CoT), which applies only CoT prompting, achieves intermediate results, suggesting CoT contributes positively but benefits from being combined with few-shot and prompt sampling.

\item The similar or slightly better results on the test set compared to evaluation indicate stable training and reasonable generalization. The relatively lower EM scores compared to PM and SM highlight the challenge of exact docid generation, leaving room for future improvements.
\end{enumerate*}

\subsection{Training efficiency}

\begin{table}[t]
\vspace{-3mm}
\caption{GPU memory usage and training efficiency of different models.}
\label{tab:gpu_mem_train_eff}
\centering
\begin{tabular}{l@{}cc ccc}
\toprule
\textbf{Model} & \multicolumn{2}{c}{\textbf{GPU Memory (MiB)}} & \multicolumn{3}{c}{\textbf{Training}} \\
\cmidrule(r){2-3}
\cmidrule{4-6}
 & Model only & Training (median) & Epochs & Time & Samples (mean) \\
\midrule
ReasonGR (Zero) & 653.875 & \phantom{0}8496.19 & 40 & 03:36:38 & 267.881 \\
ReasonGR (CoT)  & 653.875 & 25424.19           & 60 & 06:49:29 & 141.720 \\
ReasonGR        & 653.875 & 18280.19           & 55 & 05:11:09 & 186.506 \\

\bottomrule
\end{tabular}
\vspace{-4mm}
\end{table}

\paragraph{Memory usage}
Table~\ref{tab:gpu_mem_train_eff} shows that the baseline model memory is similar across ReasonGR variants. Training memory varies notably: ReasonGR (Zero) uses about 8.5 GiB, ReasonGR with CoT prompting uses the most, and full ReasonGR is in between. The higher memory for CoT variants likely comes from longer or more complex input and output sequences.

\paragraph{Training time}
ReasonGR (Zero) trains 40 epochs in roughly 3 hours, processing 268 samples/sec. ReasonGR (CoT) takes nearly 7 hours for 60 epochs with 142 samples/sec throughput. Full ReasonGR trains for about 5 hours 11 minutes over 55 epochs at 187 samples/sec. CoT prompting adds computational cost, increasing training time and reducing throughput, while the Zero variant is more efficient but less reasoning-capable.

\section{Discussion} \label{ch:discussion}

The few-shot model with LoRA performs best overall, with the CoT model also outperforming the baseline, showing that prompting improves generative retrieval. However, limitations include: (i) the dataset was adapted from non-IR tasks, which may affect generalizability; (ii) the 512-token input limit is shorter than the average document length, and the FLAN-T5-base model may be too small for effective CoT reasoning \cite{wei2022chain}, future work could explore larger or decoder-only models or add a classification head to separate reasoning from retrieval; and (iii) using sample-level loss may reduce fine-grained feedback, so token-level loss could better focus on relevant tokens and reduce noise from reasoning text.

\section{Conclusion} \label{ch:conclusion}

We have presented ReasonGR, a GR framework that enhances reasoning ability through guided prompting and efficient fine-tuning. By combining few-shot and CoT prompting with LoRA and an adaptive loss, ReasonGR improves retrieval accuracy on complex numerical reasoning tasks, shown on the FinQA dataset. Our results demonstrate that reasoning-enhanced prompting boosts performance while keeping training efficient. Though limited by model size and dataset scope, this work shows the promise of reasoning-driven generative retrieval. Future work will explore larger models, finer loss designs, and broader evaluation benchmarks.

\subsubsection{Acknowledgements.}

This preprint has not undergone peer review (when applicable) or any post-submission improvements or corrections. The Version of Record of this contribution is published in Advances in Information Retrieval, 48th European Conference on Information Retrieval, ECIR 2026, and is available online at \url{https://doi.org/10.1007/978-3-032-21300-6_17}.
This work is funded by Ahold Delhaize, through AIRLab Amsterdam, by the Dutch Research Council (NWO), under the following project numbers 024.004.022, NWA.1389.20.183, and KICH3.LTP.20.006, and by the European Union under grant agreements no.\ 101070212 and 101201510.
We thank SURF for support in using the Dutch national supercomputer Snellius.
All content represents the opinion of the authors, which is not necessarily shared or endorsed by their respective employers and/or sponsors.

\subsubsection{Disclosure of interests.}

The authors have no competing interests to declare that are relevant to the content of this article.

\bibliographystyle{splncs04}
\bibliography{bibliography}

\end{document}